\documentclass[12pt]{article}
\usepackage{amssymb,amsmath}
\usepackage[dvips,unicode]{hyperref}
\usepackage[dvips]{graphicx}

\begin{document}
 \title {\bf  \large  On  neutron-antineutron oscillation processes in solar
  cosmic-rays and  the possibility of their observation}
\author { L. S. Molchatsky \thanks{Department of Theoretical Physics,
 PGSGA,  M.~Gor'ky Street 65/67, 443099 Samara, Russian Federation}
\thanks{ e-mail:\,levmol@yandex.ru}}

\date{}
\maketitle

\begin{abstract} \noindent
 The experimental observation of  neutron - antineutron oscillations  is 
 one of the oldest  problems of elementary-particle  physics.  However, 
 the search for these processes  in the  laboratories on the Earth
 have thus far yielded no results. In this paper, we consider the possibility 
 of finding $n - \tilde n$ oscillations in the solar cosmic rays  at flare 
 increases. There are two arguments in favour of such an approach 
 to this problem: a long distance for a neutron run and a weak magnetic 
 field in the solar cosmic-ray environment. Therefore, the presence of  
 antinucleons in solar cosmic rays may be a evidence of the existence 
 of $n - \tilde n$ oscillations. The intensities expected of the $\tilde n$ and 
 $\tilde p$  fluxes near the Earth are found.   
\end{abstract}


\section*{\normalsize 1. \, Introduction}
 The notion of  baryon number (or baryon charge)  was introduced by 
 Stueckelberg in 1938 to explain the stability of atoms.  
 However, numerous experiments give evidence to the fact that 
 a baryon field is absent. In view of the contemporary physical theories, 
 based on  symmetry principles, it means that  baryon charge does not 
 make dynamical sense and, consequently, it is not related  to any 
gauge symmetry. In turn, it means that baryon number can not be 
strictly   reserved. 

 There are only two kinds of  possible experiments aimed on  the
 observation of the processes with the non-conservation of  
 baryon number: attempts to reveal  proton decays and  
  neutron-antineutron oscillations.

 Experiments on  proton decays determined the lower limit for  mean 
 lifetime  of this particle: $\tau_p > 10^{33}$ years [1]. The possibilities of 
 these experiments fall off. The experiments, aimed at finding $n - \tilde n$  
 oscillations, are carried out in  Grenoble. In these experiments, 
 neutrons travel from the reactor  to the target in which  antineutrons 
 should be annihilated.  The lower limit for the oscillation period 
is $T_0  >  0.86 \times 10^8 \, s$ [1].  At present, the main question 
discussed is the experiments with  ultra-cold  neutrons confined in 
 a trap [2, 3]. 

 We consider here the possibility for finding the $n - \tilde n$   oscillations 
 in the solar cosmic-ray at flare increases. According to the astrophysical 
 data [4 - 6], powerful neutron fluxes arise at  solar flare increases. 
 These fluxes were detected by many  neutron monitors around the world. 
 This fact is very attractive taking into account that the suppression 
 of the $n- \tilde n$  oscillations in cosmic-rays should be weaker 
 than  that on the Earth.

 \section*{\normalsize  2. \, The $n - \tilde n$ mixing. Oscillations}
 At a non-conservation of baryon number, the states $|n>$ and $|\tilde n>$ 
 are not stationary. In this case the $|n>$ and $|\tilde n>$ states 
 are related  with stationary ones of  $|1>$ and $|2>$ by the orthogonal 
 transformations
 \begin{eqnarray} 
 |1>  & = & cos\theta \, |n> + sin\theta \, |\tilde n>, \nonumber \\ 
 |2>  & = & -sin\theta \, |n> +cos\theta \, |\tilde n>, 
 \end{eqnarray}
  where $\theta$ is the mixing angle which defines the power of  breakdown 
 of a charge conservation law.

Let an initial state of the neutron-antineutron system at $t' = 0$  be neutron, 
i. e. $\psi (0) = |n>$. If the transformation (1) takes place, then, as a result 
of   quantum  mechanical evolution, at $t' = t > 0$ the system should be 
 in the state 
   \begin{multline*}
        \psi( t ) = cos \theta \, e^{-i E'_1 t}  |1> - sin \theta \, e^{-i E'_2 t} |2>  
         = cos\theta \, e^{-i E'_1 t} (cos\theta  \, |n> + sin\theta \, |\tilde n>)  \\
        - sin\theta \, e^{-i E'_2 t} (-sin\theta \,|n> +cos\theta \, |\tilde n>)  
        = (cos^2 \theta  \, e^{-i E'_1 t} + sin^2 \theta \, e^{-i E'_2 t}) \, |n>  \\
        +  (1/2) sin 2\theta \,(e^{-i E'_1 t} - e^{-i E'_2 t})  |\tilde n>. 
 \end{multline*}
 From this equation we find that the amplitude for transition $n \to \tilde n$ 
   \, is 
   \begin{equation} 
 A(n \to \tilde n) =  \frac 1 2 sin 2\theta \, ( e^{-i E'_1 t} - 
 e^{-i E'_2 t}) .
 \end{equation} 
 Here and further we suppose $\hbar = 1$. 
 
 At conservation of  baryon number , the energetic levels of a free  system
 are generated , i.e. $E'_1 = E'_2$ . In other cases, the stationary state levels 
 are splitting,  i. e. $\Delta E' = E'_1 - E'_2 \ne  0$.  For those cases, we 
 write  the energetic levels  in the form 
$$
 E'_1 = E' + \frac{\Delta  E'}{2}; \quad  
 E'_2 = E' - \frac{\Delta  E'}{2}.
$$  
 Besides,  it is necessary to take into consideration the fact that free 
 neutrons and antineutrons are  the $\beta$- decay particles and 
 therefore energetic levels of the system have  the  width 
 $\Gamma = 1/\tau_n$, where $\tau_n$ is mean lifetime for 
 neutron.  To take into account  neutron instability we assume 
 $E' = E - i \Gamma/2$. As consequence of these additions, formula (2) 
 acquires  the form 
 $$ 
   A(n \to \tilde n) = - i \, e^{- i E t} \, e^{- \Gamma t /2} \, sin{2 \theta} \, 
   sin{(\Delta E /2)t}.
 $$ 

 Thus, the probability for the conversion $n - \tilde n$ is defined by 
      \begin{equation}
    P(n \to \tilde n) = e^{-\Gamma t} \, sin^{2}{2\theta} \, 
    sin^2 {(\Delta E /2) t}.
 \end{equation}  

  \section*{\normalsize 3. \, Influence of external field upon  $n - \tilde n$ 
 oscillations } 
 At  baryon charge non-conservation, the off-diagonal matrix element 
 $\delta $ of the Hamilton operator in the charge representation is not 
 equal to zero, i. e. 
 $$ 
  \delta  = <\tilde n|\hat H|n> = <n|\hat H|\tilde n> \ne 0 . 
 $$
 The parameter $\delta$, as well as the angle $\theta$, point to degree 
  of the  baryon charge non-conservation. Their relation is given by 
  \begin{equation}
 sin {2\theta} = \frac{2\delta}{E_1 - E_2}, 
   \end{equation} 
  where 
   \begin{equation}
 \Delta E =  E_1 - E_2 = \sqrt{(E_n - E_{\tilde n})^2 + 4 \delta^2}. 
   \end{equation} 
  Here $E_n = <n|\bar H|n>, \,  E_{\tilde n} = <\tilde n|\bar H|\tilde n>$
 are the diagonal matrix elements of the Hamilton operator. 
 In deducing formulae (4) and (5) transformation (1) has been used 
  in inverse form. 

 So from formulae (3), (4), and (5) we extract the following conclusion. 
 Free oscillation is described by 
 \begin{equation}
 P_{free} (n \to \tilde n) = e^{-\Gamma t} \,sin^2 {\delta  t}, 
 \end{equation}
 because in this case $E_n = E_{\tilde n}.$ 
 
 Difference between $E_n$ and $E_{\tilde n}$ due to magnetic field is 
  $| E_n  -  E_ {\tilde n} | = 2 |\mu| \, B $, consequently, 
   \begin{equation}
 P_{field} (n \to \tilde n) = \frac{\delta^2}{(\mu B)^2 + \delta^2} \, 
 e^{-\Gamma t}   \, sin^2 {\sqrt{(\mu B)^2 + \delta^2} \,  t}. 
 \end{equation}

 \section*{\normalsize  4. \, Antineutrons and antiprotons  in  vicinity 
  of the   Earth as a consequence of  $n-\tilde n$ ~ oscillations}
 If  the oscillation exists, then the transitions 
   \begin{equation} 
 n \to \tilde n \to \tilde p + e^+ + \nu_e 
  \end{equation}
 must occur in  solar neutron fluxes. Therefore, the observation 
 of    the  $\tilde n$ and  $\tilde p$ particles in  solar cosmic-ray 
 might give evidence to the existence of these processes. 

 We can now proceed to the estimate of count rate 
 of the $\tilde n$ and  $\tilde p$ fluxes near the Earth with respect to 
 the $n$ flux at the Sun.

 Proceeding from the experiment data [1], period free oscillation 
 is assumed to be $T_0  \approx    0.86 \times 10^8 \, s$. By using  this 
 we find corresponding value for the mixing parameter: 
 $$ 
 \delta \approx 1 \times 10^{-23}\, eV .
 $$  
 It is deduced from (6).

 As to magnitude $\mu B$, its estimate gives  
 $\mu B \approx 3 \times 10^{-16} \, eV$  because
$B \approx 5 \times 10^{-5} \, G$  in the solar cosmic-rays. 
 So we establish that $(\mu B)^2 \gg \delta^2$.

 In addition to this, we have $T  \ll t$, where $T$ is the oscillation period  
  in interplanetary space and $t$ is  run time for particles.  This 
 relation takes place because according to (7) the oscillation period 
 in magnetic  field is $T =\pi/(\mu B) \approx  7  s$, whereas   the run time  
 for  the neutron  with  $E =100 \, MeV$  is  $t = 1 \times 10^3 \,  s $. 

 For these conditions, formula (7) takes the form
  \begin{equation}
 P(n - \tilde n) = \frac{\delta^2}{2  (\mu B)^2} \,  e^{- \Gamma t}.   
 \end{equation}	
 This formula defines the $n-\tilde n$  transition probability at a neutron 
 run from the Sun to the Earth. 

 As a consequence of this, the probability of the $\tilde p$ production 
 by means of transitions (8) is  
  \begin{equation}
 P(n \to \tilde n \to \tilde p e^+ \nu_e) = 
 \int\limits_0^t  P_{n - \tilde n}(t') \Gamma \, dt' = 
  \frac{\delta^2}{2   (\mu B)^2} \, (1 - e^{- \Gamma t}).	
 \end{equation}

 Formulae (9) and (10) show that a magnetic field leads to a strong 
  suppression of the  $n-\tilde n$ oscillations. Intensities of these 
processes    on the Earth surface and in the environment of the solar 
 cosmic-rays are  related by 
$$ 
  I_E/I_{c- r} =  (B_{c-r}/B_E)^2 \propto 10^{-8}.  
 $$
 This result is the main evidence in favour of the experiments with  solar 
  cosmic-ray.

The results of calculations for the  $\tilde n$ and $\tilde p$ fluxes 
expected   at the Earth is summarized in Table 1. 

 \smallskip  
 \begin{center}
\begin{tabular}{|c|c|c|c|} 
 \multicolumn{4}{c}{\bf Table 1 .  \it The $\tilde n$ and $\tilde p$ fluxes 
 near the Earth }\\ 
   \hline 
 E, MeV  & $ \frac{ I(\tilde n)}{I(n)}$  & $\frac{I( \tilde p)}{I(n)}$  & 
 $\frac{ I(\tilde n) + I(\tilde p)}{I(n)}$ \\ 
 \hline 
 10        & $  0.1 \times 10^{- 16}$  & $ 6.0 \times 10^{- 16} $  & $ 6.1 
\times 10^{- 16}$ \\
 100     &  $1.6 \times 10^{- 16}$  &  $3.5 \times 10^{- 16} $ &  $5.1 
  \times 10^{- 16}$ \\
 !000    &  $2.2 \times 10^{- 16}$  & $ 0.9 \times 10^{- 16}$  & $ 3.1 
\times 10^{- 16}$ \\  \hline 
\end{tabular} \\
 \end{center}
  Here $ I(\tilde n) / I(n)$ and $ I(\tilde p) / I(n)$  are expected ratios 
 of the  $\tilde n$ and $\tilde p$ fluxes at the Earth to  initial neutron 
flux  from the Sun. The calculations are carried out by means of formulae 
  (9) and (10). 
  
 The results obtained show that the $\tilde p$ flux should be dominating
 in the antinucleon flux at  low-energies (about $10 \, MeV$).  However, 
  the   $\tilde n$  rate  increases as the energy increases. At the energies 
 about $100 \, MeV$  the fluxes of   $\tilde n$  and $\tilde p$ should be 
 approximately  equal. Nevertheless, the total antinucleon flux decreases 
 with increasing  energy.

 As to the neutron flux emanated from the Sun, it falls 
 as $I \propto E^{- 5.5}$   [4 ].  

 Thus, the existence of antinucleons in the solar cosmic-rays near the Earth  
 should be dominating at low-energies.

  \section*{\normalsize 5. \, Concluding remark} 
 It should be noted that neutrino oscillations have been found just in 
 the neutrino fluxes traveling from the Sun to the Earth.   It is possible that 
 the search for neutron-antineutron oscillations in solar neutron fluxes 
 may also be effective.

 
 \bigskip 

 \noindent {\bf \normalsize References}\\  
 \begin{itemize}
 \item[ [1]] W.-M. Yao et. al. (Particle Data Group),  J. Phys. G {\bf  33},
       1 (2006).
 \item [[2]] B. O. Kerbikov, A. E. Kudryavtsev, V. A. Lensky, 
       Zh. Eksp. Teor.Fiz.        {\bf 125}, 476 (2004).
 \item[[3]]  V.  Ignatovich,  Phys. Rev. D {\bf  67}, 016004 (2003).  
 \item[[4]]  H. Debrunner, J. A. Lockwood, J. M. Ryan,  The Astrophysical 
  Journal   {\bf 409}, 822 (1993). 
 \item[[5]]  H. Debrunner et. al.,  The Astrophysical Journal {\bf 479}, 997 
     (1997). 
  \item[ [6]]  Yu. D. Kotov,  Usp. Fiz. Nauk {\bf 180}, 647 (2010).
 \end{itemize}
   
\end{document}